\documentclass[aps,letterpaper,pra,twocolumn,showpacs]{revtex4}
\usepackage{graphicx}
\usepackage{amsfonts}
\usepackage{amsmath}
\usepackage{amssymb}
\usepackage{color}

\renewcommand{\vec}[1]{{\mathbf #1}}

\begin{document}

\title{Direct laser writing of three dimensional network structures as templates for disordered photonic materials.}

\author{Jakub Haberko$^{1,2}$ Nicolas Muller$^{1}$ and Frank Scheffold$^{1,\ast}$}

\address{$^{(1)}$Physics Department and Fribourg Center for Nanomaterials, University of Fribourg, Chemin de Mus\'{e}e 3, 1700 Fribourg, Switzerland}
\address{$^{(2)}$Faculty of Physics and Applied Computer Science, AGH University of Science and Technology, al. Mickiewicza 30, 30-059 Krakow, Poland }

\date{\today}

\begin{abstract}  In the present article we substantially expand on our recent study about the fabrication of mesoscale polymeric templates of disordered photonic network materials \cite{Haberko13}. We  present a detailed analysis and discussion of important technical aspects related to the fabrication and characterization of these fascinating materials. Compared to our initial report we were able to reduce the typical structural length scale of the seed pattern from $a=3.3\mu$m to $a=2\mu$m, bringing it closer to the technologically relevant  fiber-optic communications wavelength range around $\lambda \sim 1.5 \mu$m. We have employed scanning electron microscopy coupled with focused ion beam cutting to look inside the bulk of the samples of different height. Moreover we demonstrate the use of  laser scanning confocal microscopy to assess the real space structure of the samples fabricated by direct laser writing. We address in detail question about scalability, finite size effects and geometrical distortions. We also study the effect of the lithographic voxel shape, that is the ellipsoidal shape of the laser pen used in the fabrication process. To this end we employ detailed numerical modelling of the scattering function using a discrete dipole approximation scheme. 
\end{abstract}


\maketitle
\section{Introduction}
\label{intro}

Structured dielectric materials in three dimensions can exhibit photonic properties that allow to control the propagation of light \cite{Haberko13,Joann,Yab,ref10,ref11,Soukoulis94,Soukoulis2011}. For crystalline structures, a complete or incomplete photonic band gap emerges and the propagation of light is hindered or even completely suppressed over a certain range of wavelengths. Full photonic band gaps open up for dielectric materials with a sufficiently high refractive index contrast \cite{Soukoulis94}. The change in transport properties is accompanied by a reduction in the local density of state which results in increased lifetimes for embedded light emitters such as fluorescent molecules \cite{Lodahl2004}. Interestingly, it appears that many of these unique properties are not tied exclusively to crystalline structures  \cite{Torq09,ref2,Rec11,ref13,ref3,ref4,ref5,ref5b,ref5c}. In a recent numerical study Florescu et al. demonstrated that particular designer disordered materials can display large, complete photonic band gaps in two dimensions  \cite{Torq09}. Mapping hyperuniform point patterns with short-range geometric order into tessellations allows the design of interconnected networks that gives rise to enhanced photonic properties. A point pattern is \emph{hyperuniform} whenever infinite wavelength density fluctuations vanish \cite{Torq03}. First experimental data obtained for such disordered 2D structures in the microwave regime have been reported post-submission of the present work in \cite{Man10}. The numerical study by Florescu et al. has recently been extended to three dimensions by Liew et al. \cite{Liew}. It was found that for a  refractive index $n >3$ (in air) the optical density of states is  significantly reduced at certain wavelengths and a photonic band gap opens up \cite{Liew}. 
\newline \indent Although some of these concepts, both for crystalline and disordered materials, can be tested on macroscopic length scales using microwaves \cite{microwavePC} or ultrasound \cite{ultrasoundPC}, most of the fundamental interest and possible technological applications concentrate on the infrared or visible range of the optical spectrum \cite{Joann,ref12,Soukoulis2011,FibCom}. This means that typical structural length scales of the materials need to be scaled down to the submicron range.  Periodic photonic structures can be obtained by self-assembly of micro- and nanosized polymer or silica (SiO$_2$) spheres resulting in (pseudo) band gaps that can be tuned easily from visible to infrared wavelengths \cite{ref11,siliconPCcolloids}.  Creating complex three dimensional disordered structures, such as photonic crystals with integrated waveguides \cite{PCwaveguides11} or random materials like the ones suggested by Florescu et al. \cite{Torq09}, however mandates the use of more sophisticated \emph{top-down} fabrication tools. Classical lithography allows sub-$100$nm microfabrication but the method is restricted to 2D. In stereo-lithography, used in modern 3D printers, thin layers of a photopolymers are exposed according to a sliced representation of the desired structure. Due to the layer-by-layer approach the axial resolution is however limited to several micrometers even for the most advanced realizations of this technique \cite{Stereolitho}. Submicron scale periodic patterns, such as photonic crystals, can be produced with multi-beam interference lithography \cite{MBIL1} and proximity field nanopatterning \cite{PnP}. Probably the most versatile high resolution 3D lithography however is two-photon direct laser writing (DLW)\cite{PCwaveguides11,ref12,Maruo08}. Here a polymer photoresist absorbs two infrared photons simulateously and the combined energy leads to solidification of the resist. Since the probability for two photon absorption scales quadratic with laser intensity the process is highly selective both laterally and axially providing submicron resolution. The technique has been extended lately towards sub-diffraction limited resolution \cite{FischerWegener13} imitating the concept of stimulated emission depletion(STED) known from super resolution scanning fluorescence microscopy \cite{Hell94,Hell99}. 
\newline \indent  Recently we reported on the first successful DLW - fabrication of mesoscale polymeric templates of the disordered dielectric materials suggested by Florescu et al. \cite{Torq09}. We were able to fabricate hyperuniform network structures with features on submicrometer length scales using direct laser writing into a photoresist \cite{Haberko13,Wegener12}. Although the feasibility of  fabricating these interesting network structures has been shown, further progress towards the realization of actual photonic materials mandates a better and more detailed understanding of the many aspects that control the material properties and ultimately their optical response. In the present article we substantially expand on our previous study and present a detailed analysis and discussion of some of the most important questions related to the fabrication and characterization of these interesting materials. We employ scanning electron microscopy coupled with focused ion beam cutting to look inside the bulk of the samples. For the routine non-destructive 3D visualization of the structures we use in-house laser scanning confocal microscopy. Compared to our initial study\cite{Haberko13}, we have succeeded in reducing all relevant length scales by a factor 5/3 while keeping the quality of the structures unchanged. Moreover we examine the role of geometric distortions and finite size effects. We study the effect of the lithographic voxel shape, that is the ellipsoidal shape of the laser pen used in the fabrication process as well as macroscopic distortions that emerge as a result of internal stresses in the polymer network during fabrication and development. To this end we analyze in detail numerical calculations of the diffraction patterns of weakly scattering structures using a discrete dipole approximation. 

\begin{figure}\centering
\resizebox{0.4\textwidth}{!}{\includegraphics{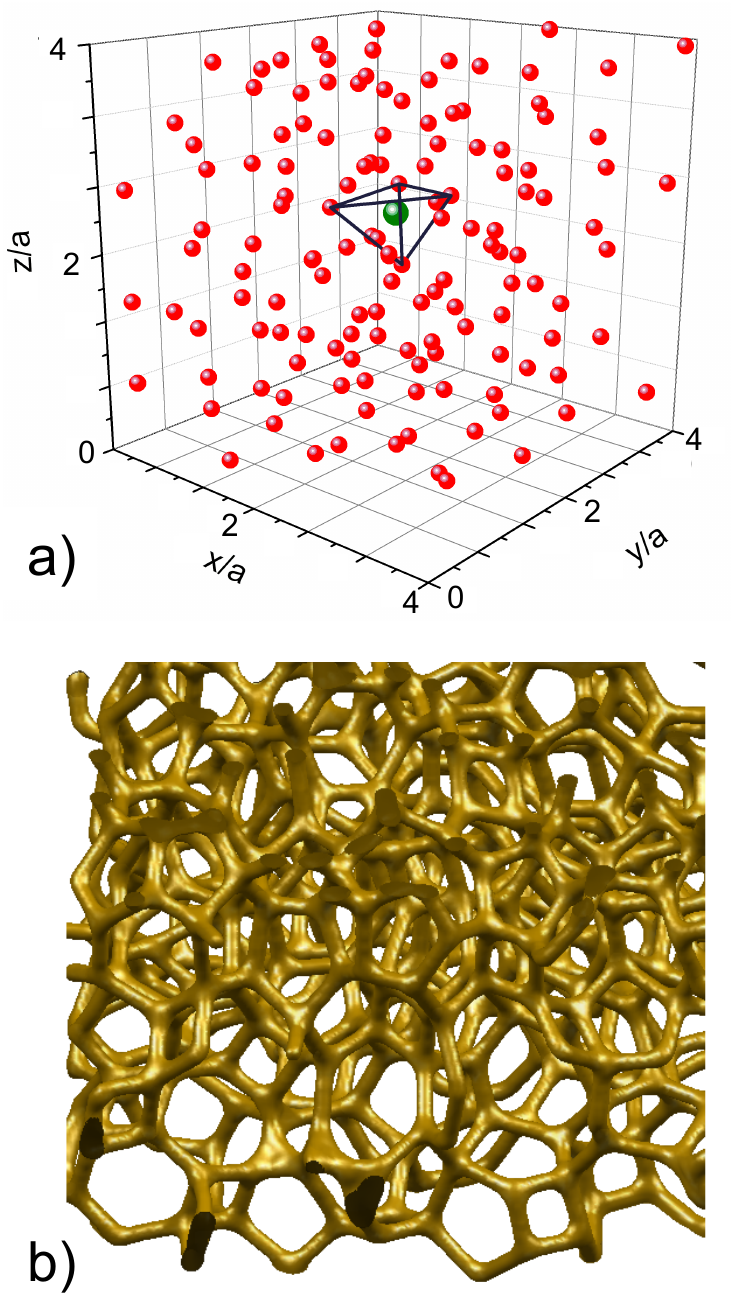}
} 
\caption{(Color online). Design of amorphous photonic network structures.  (a) the small spheres represent the seed point pattern derived from a jammed assembly of hard spheres with diameter $a$ \cite{ref15}. The dark lines show an example for the Delauny tesselation of the seed pattern resulting in a network of tetrahedron. The larger dark sphere shows the centroid position of the tetrahedron. (b) The three dimensional hyperuniform network obtained by connecting the centroid position of neighboring tetrahedrons and replacing the connection lines with solid rods. 
}
\label{Fig1}      
\end{figure}

\section{Fabrication of mesoscale polymeric templates by direct laser writing}
\label{sec:2}
\subsection{Implementation of the laser writing protocol} 
\begin{figure}
\resizebox{0.5\textwidth}{!}{\includegraphics{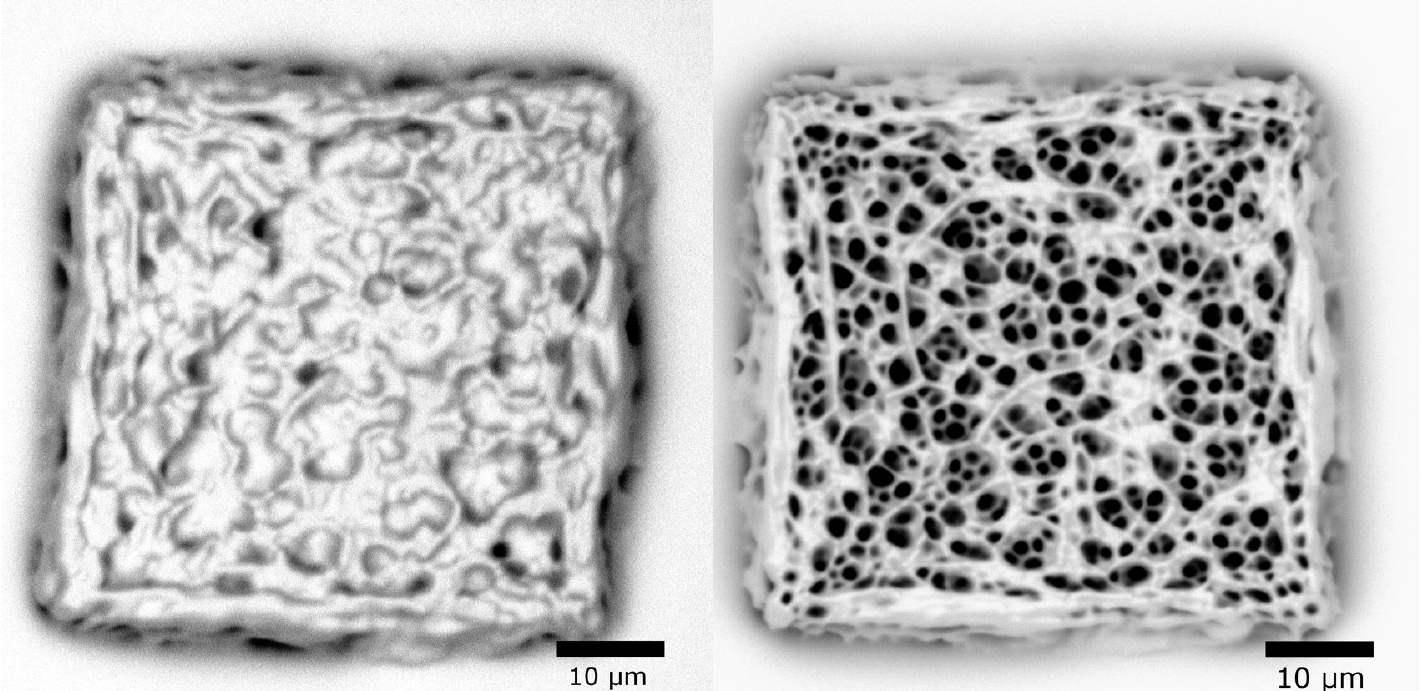}
} 
\caption{Electron micrographs of some initial trials to fabricate polymeric network structures by direct laser writing (DLW) illustrating the collapse and deformation of the structures due to an inadequate writing protocol.
}
\label{Fig2}    
\end{figure}

Mesoscale polymer network structures can be fabricated by direct laser writing (DLW) using a two-photon polymerization process. Here we fabricate polymeric network structures using the Photonics Professional direct laser writing platform (Nanoscribe GmbH, Germany) \cite{Haberko13}.  For the fabrication of hyperuniform network structures, we implement a protocol derived by Florescu and coworkers to map a hyperuniform point pattern into tessellations for a photonic material \cite{Torq09,Liew}. As a starting point we use the centroid positions from a maximally randomly jammed assembly of spheres of diameter $a = 2-3.3\mu$m  with a volume filling fraction of $\phi \simeq 0.64$ \cite{ref15}.  As shown by Torquato and coworkers \cite{Zac11} such seed structures indeed possess the required hyperuniform properties \cite{stealthy}. Next we perform a 3D Delaunay tessellation of the seed pattern: tetrahedrons are constructed in such a way that no sphere center is contained in the circumsphere of any tetrahedron in the tessellation. The centers-of-mass of neighboring tetrahedrons are then connected resulting in a 3D random tetrahedral network, Figure \ref{Fig1}. The network structure connection lines are written into a IPG-780 negative tone photoresist (Nanoscribe, Germany). The laser writing pen has an elliptical shape with a typical dimension for the long axis of  $800$nm and $300$ nm for the short axis. In order to expose the photoresist the sample is moved with a three axes piezo-stage while the position of the pen remains unchanged (stage-scannig configuration). In principle the (static) positioning accuracy of the piezo scanning stage (and thus the pen) is specified to less than five nanometers. However, as explained below, rapid movement over long distances can still adversely affect the precise positioning of the laser pen in the resist. The overall size of the pen can be adjusted, within a certain range, by adapating the power or residence time of the fs-laser focal spot \cite{shadedring2}. Thus the volume-filling fraction of the rods can be controlled by the writing speed or exposure time of the photoresist. In our case the lower acceptable limit is set by mechanical network failure while the upper limit is due to the fact that overexposure will lead to a collapse of the network structure by the fusion of neighbouring rods, Figure \ref{Fig2}.   The aspect ratio of the pen is only weakly affected by the power setting. Both the approximate size and the aspect ratio are determined by the point spread function of the illuminating microscope objective which  determines the isointensity surface. Above a certain threshold intensity the polymer photoresist is cross linked. The point-spread-function in turn is set by the refractive index of about $1.52$ of the material and the numerical aperture of the objective, typically $1.4$. In the past it was shown that placing a shaded ring filter in the beam path can reduce the aspect ratio of the writing pen from $\sim 2.7$ to $\sim1.8$  \cite{shadedring1,shadedring2}. For fabricating the structures with $a=3.3\mu$m \cite{Haberko13}, we have used the standard configuration of the DLW system using an oil immersion objective with a numerical aperture 1.4. More recently, for structures featuring a reduction of length scales down to $a=2\mu$m, we use dip-in direct laser writing \cite{Dip,Freyman10} using a shaded ring filter \cite{dip shaded}. 

\begin{figure}\centering
\resizebox{0.48\textwidth}{!}{\includegraphics{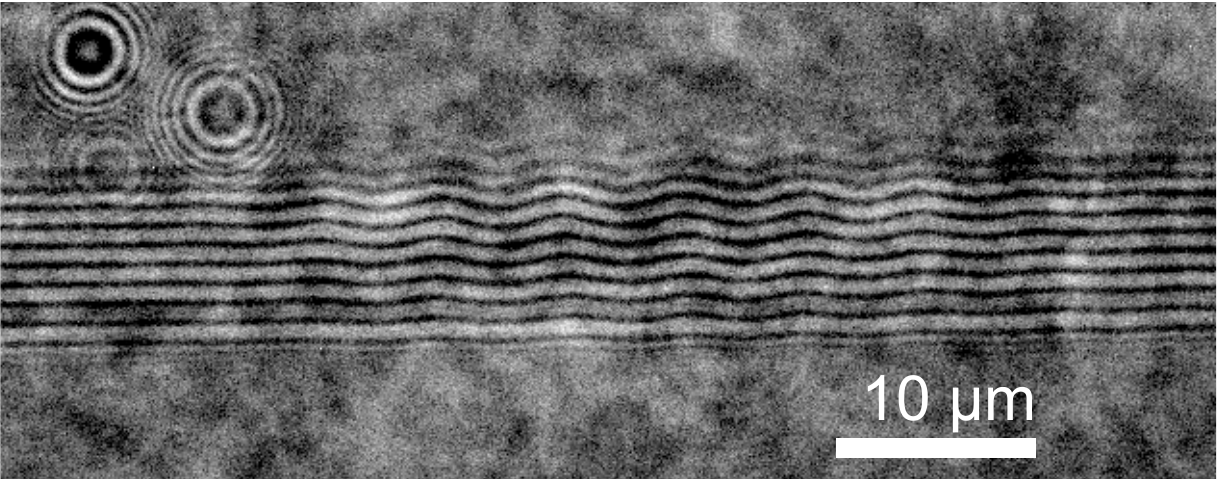}
} 
\caption{Light microscopy image taken directly after writing several parallel lines of length 100 $\mu m$ from top to bottom.  Writing each line takes 1 second (at 100 $\mu m$ /s writing speed) and the wait time between the different lines is 0.2 sec. Within seconds after exposure of the photoresist the lines become wiggly and deformed.}
\label{Fig3}      
\end{figure}
In comparison to the well-established case of periodic rod assemblies  \cite{ref12} the fabrication of these designer disordered structures is very demanding and requires optimization of the writing protocol. To illustrate the complexity of fabricating these structures we show in Figure \ref{Fig2} a selection of electron micrographs of our failed initial trials. Particular care must be taken how to set up the writing protocol since there are no obvious rules how to write a random free-standing network structure at optimal resolution. Moreover it must be ensured that the structure remains mechanically stable in the soft gel photoresist IPG-780 throughout the writing process of about 1 h. As we found out in the course of this work, the difficulties of writing random 3D network structures are mainly due to the hysteresis of the piezoelectric stages and due to problems with unconnected rods. If the stage were to travel rapidly a long distance between two consecutively written rods, this long jump can, presumably due to the inertia of the sample stage, lead to a slightly inaccurate positioning of the second rod. The architecture of the network required that rods must be connected in such a way that in each vertex point four of them meet, Figure \ref{Fig1}. Inaccurate positioning means that some of the rods have dangling ends rather than being connected to a proper vertex point or worse would not be part of the continuous network altogether. Apart from the deviations of the designed structural properties, this would also adversely affect the network$'$s mechanical stability. Therefore, the writing scheme should favor short distances between consecutively written rods (preferably writing continuous broken lines, whenever possible). Secondly, if an unconnected rod is written in the bulk of the photoresist, it quickly undergoes severe distortions, Figure  \ref{Fig3}. Within seconds after exposure such lines become wiggly and as a result move away from their initial positions. We have been unable to decipher the exact reason for this behavior, but we assume it may result from mechanical stress induced by exposing the photoresist to the writing laser. While this is not a problem for periodically ordered layered structures  it becomes truly challenging when fabricating complex disordered networks.  Hence, during the writing process each rod should be connected to the substrate or to another rod as quickly as possible. 
\newline \indent To overcome these difficulties we divide the writing process into writing subregions. To this end, we select stacks of cubic volumes of side length roughly 1.5$\times a$ and sorted our rods in the following way: i) once all rods belonging to a certain cube are written, we proceed to a neighboring cube, ii) first all cubes closest to the substrate are filled with rods, then the ones lying higher above and so on until the whole network has been written. This procedure solves both the problem of avoiding long piezo movements (strictly) and the issue of connectivity (approximately). An alternative approach might be to identify continuous broken lines connecting all vertices. Among the multitude of possibilities one would then need to choose the one that ensures writing roughly from the substrate upwards. However, given the high quality of the samples already obtained by using the first protocol, we have not yet tested the feasibility of the latter approach.
\begin{figure*}\centering
\resizebox{0.7\textwidth}{!}{\includegraphics{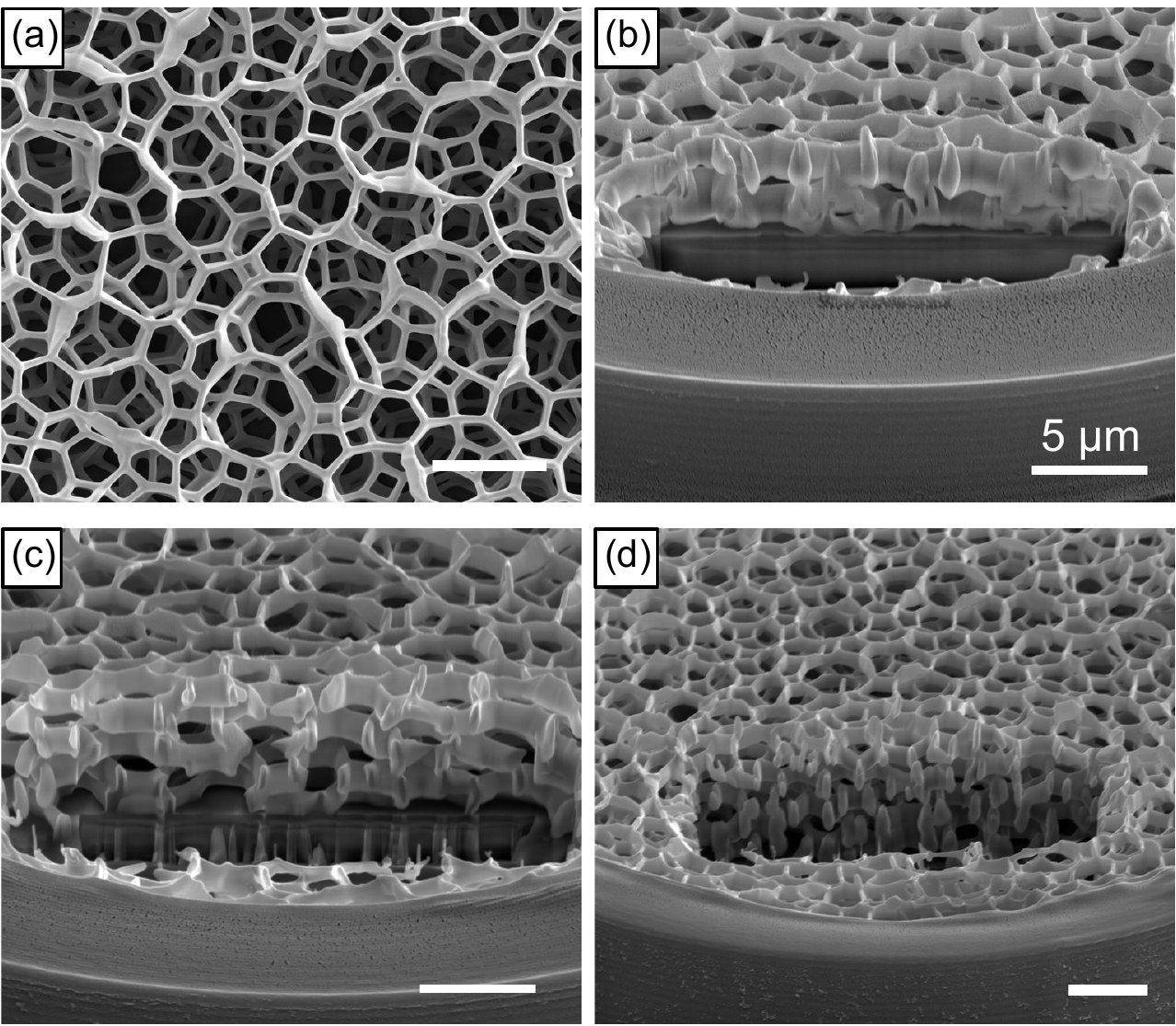}
} 
\caption{a) Scanning electron micrographs of a 3D network structure (height 8 $\mu m$), top view. b) electron micrograph of the same structure after focused-ion beam milling (oblique view), revealing the internal arrangement of dielectric rods and the glass substrate underneath.  c) oblique view of a 12 $\mu m$-high sample. d) oblique SEM view of a 16 $\mu m$-high sample. Scale bars : 5 $\mu m$.}
\label{Fig4}       
\end{figure*}
\section{Sample characterization}

\subsection{Focused ion beam millling and scanning electron microscopy}
Figure \ref{Fig4} a) displays a top-view electron micrograph of a fabricated structure, height $h=8 \mu$m. The fabrication of this structure has been first reported in \cite{Haberko13}. We use ion-beam milling in order to reveal the interior properties of our samples with heights ranging from $h=8-16 \mu$m. Figures \ref{Fig4} b)-d) show snapshots taken during the milling process. The images demonstrate that the structures are written uniformly throughout the sample volume. The images also reveal the size and shape of the ellipsoidal writing pen which we estimate to $840 \times 280$ nm$^2$. The high resolution bulk imaging allows us to estimate the polymer content in this structure to approximately $10\%$.

\begin{figure}\centering
\resizebox{0.35\textwidth}{!}{\includegraphics{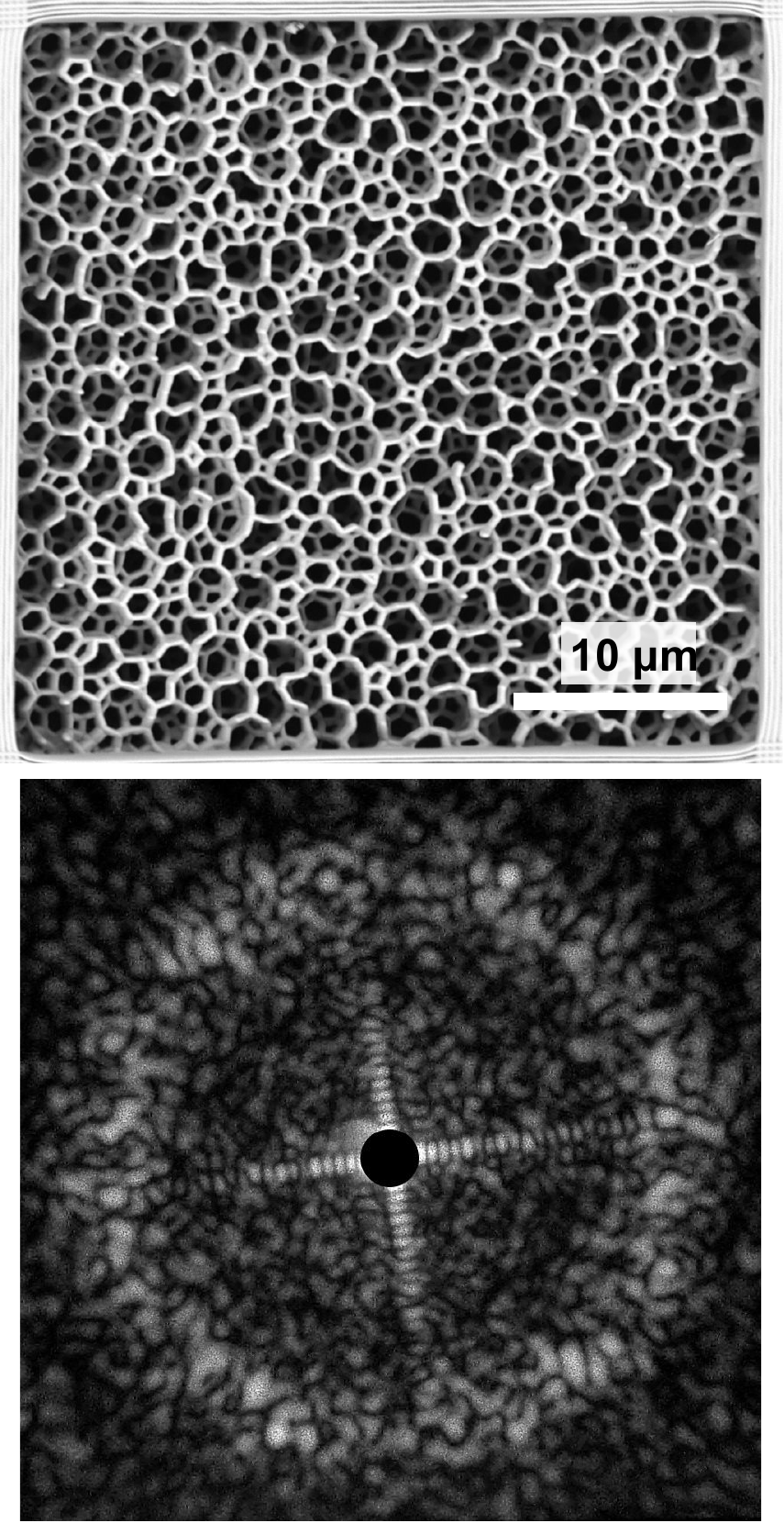}
} 
\caption{Upper panel: Scanning electron micrograph of a structure with $a=2 \mu$m. Lower panel:  Experimental light diffraction patterns of the same structure. The lateral size of the image corresponds to a scattering angle of 26 degrees (for $\lambda=633$nm). The cross shaped diffraction pattern at lower angles is due to diffraction from the sample boundaries (sample size  $40 \times 40$ $\mu m^2$. )}
\label{Fig5}      
\end{figure}

\subsection{Progressive reduction of structural length scales}
\label{sec:4}
\begin{figure}
\resizebox{0.45\textwidth}{!}{\includegraphics{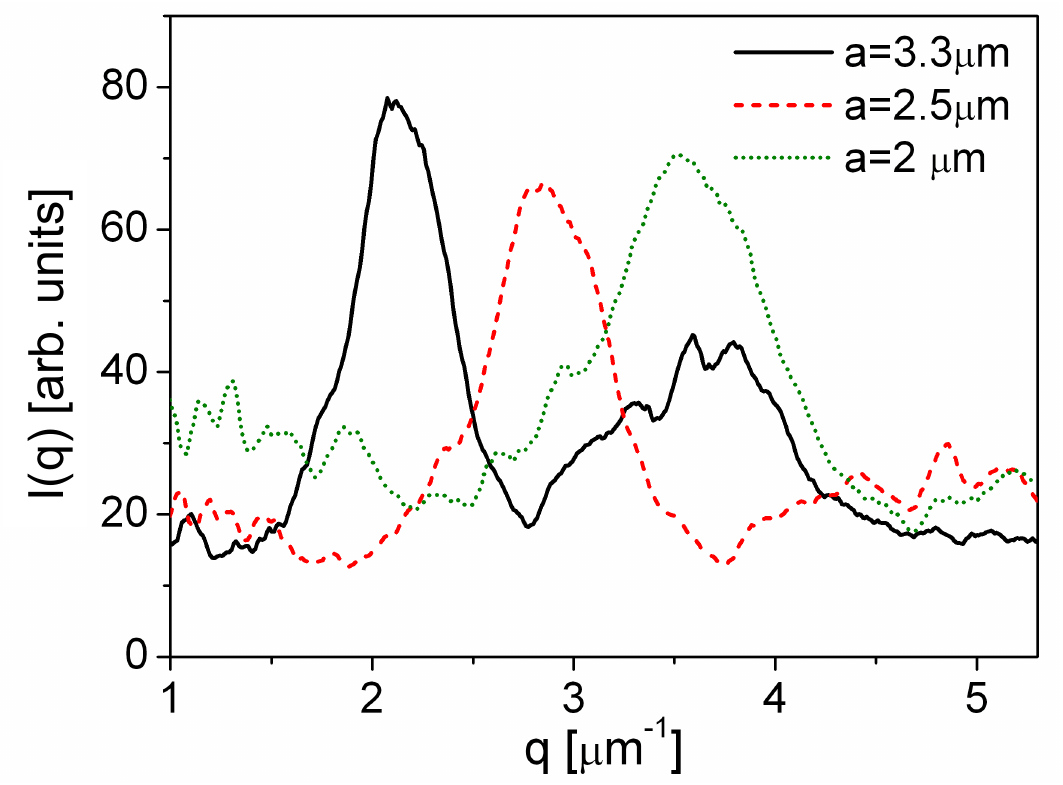}
} 
\caption{(Color online). Radially averaged experimental small angle light scattering patterns for structures with different scaling parameters $a$.}
\label{Fig6}      
\end{figure}

For the hyperuniform network structures considered here Liew et. al. \cite{Liew} predict a full band gap to open up at $\lambda \approx 1.77 a $ for $n\simeq 3.6$ and a solid filling fraction of $0.2$. We note that the photoresist used to fabricate the polymer template has a refractive index of only $n=1.52$. If transferred into a high dielectric material such as silicon (with $n\simeq 3.6$) the structures reported in our first article \cite{Haberko13}, derived from a jammed assembly of spheres with a diameter of $a = 3.3\mu$m, would feature a band gap around $\lambda \sim 5.8\mu$m. This wavelength is still about four times larger than typical telecommunication wavelengths of $\lambda \sim 1.3-1.65\mu$m \cite{ref12,FibCom}.  In order to shift the bandgap position towards shorter wavelengths all lengths need to be scaled down.  By carefully adjusting the laser writing parameters we succeeded to reduce the structural parameters of our samples by up to a factor of $5/3$ ($a = 2\mu$m) as shown in Figure \ref{Fig5}. In order to routinely assess the structural quality of the polymeric templates we perform optical diffraction experiments with a home-built small angle light scattering set-up. Experimental details concerning the set-up are described in \cite{Haberko13}. The local optical response is charaterized by the effective differential cross section d$\sigma/$d$\Omega$ or the scattering function $I(\vec{q})\sim$ d$\sigma/$d$\Omega$. In the weak scattering limit (or 1st Born approximation) the scattering function is directly connected to the structural materials properties via a simple Fourier transform $I(\vec{q}) \sim \left| {\int \rho(\vec{x}) \
e^{ -i\vec{q} \cdot \vec{x} } \
d\vec{x}} \right|^2$. Here $\rho(\vec{x})$ denotes the scattering length density which in our case is set by the refractive index contrast $\rho(\vec{x}) \sim f(\vec{x})(n/n_0-1)^2$, where $n$ is the polymer refractive index and $n_0$ the refractive index of the surrounding medium \cite{Kerker69}. Here the function $f(\vec{x})$ describes the configuration of the polymer network; i.e. $f(\vec{x})=1$ in the presence of polymer and zero otherwise.
Here we find that the scattering pattern, shown in Figure  \ref{Fig6}, remain essentially unchanged for the scaled down network structures. This indicates that reducing the length scales by up to a factor $5/3$ does not have a negative impact on the optical quality of the structures. As expected the peak value $I(q_{max})$ is observed at $q_{max}\simeq 2 \pi/a$ under index matching conditions. For a further reduction of the structural length scales we face some important challenges. The average length $d$ of dielectric rods connecting the nodes of the hyperuniform network is $d=0.39 a$ \cite{Liew}. Thus for a seed pattern with $a < 2\mu$m we would have to write rods $d<0.8\mu$m. Moreover these rods can be oriented arbitrarily in space while the elliptical writing pen has a fixed orientation. Although we still expect to be able to further reduce structural length scales, these estimates show that with the current technology it will be very difficult, if not impossible, to fabricate structures that will feature a bandgap in the range of telecommunication wavelengths. Using STED inspired DLW \cite{FischerWegener13} would in principle allow a further twofold improved resolution. However, since all of the material and optical aspects of hyperuniform photonic materials can be conveniently studied in the spectral region $\lambda = 2-5\mu$m, we are currently not exploring such further reduction in length scales until a better understanding of the properties and spectral features has been reached. Moreover other IR applications in this longer wavelength range, for example in the field of low-threshold lasing\cite{Joann,Tittel2003}, could already be targeted based on the polymeric template structures presented here.

\subsection{Three dimensional optical imaging of hyperuniform polymer network structures}
\label{sec:3}
\begin{figure*}\centering
\resizebox{1\textwidth}{!}{\includegraphics{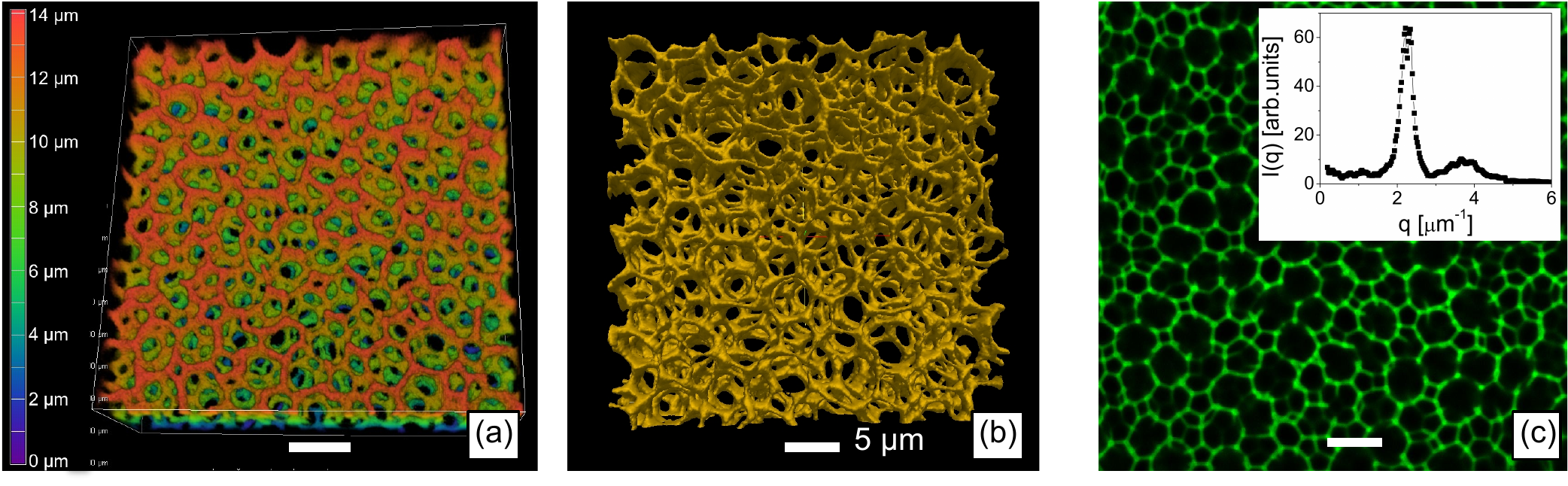}
} 
\vspace*{0.5cm}       
\caption{(Color online). Laser scanning confocal microscopy (LSCM)  images of polymer network structures with $a=3.3\mu$m.  Excitation wavelength $\lambda = 405$nm or $457$nm. a) 3D rendering of a 35x35x14 $\mu m^{3}$ volume of a sample; the axial dimnsion is given by the color code $0 \to 14\mu$m. b) isosurface rendering of a different 35x35x14 $\mu m^{3}$ region in the sample, c) a slice (44x44 $\mu m^{2}$) parallel to the sample surface at the height of approx. 7 $\mu m$, Inset: scattering function calculated from the 3D Fourier transform of the LSCM data. All scale bars are 5 $\mu m$}
\label{Fig7}       
\end{figure*}

Focussed ion beam milling combined with electron microscopy provides ultra-high resolution images of the polymer networks. The method is however destructive, somewhat time consuming and relatively expensive. Diffraction experiments on the other side allow a precise global assessment of the structure quality but are incapable of probing distortions and heterogeneities throughout the structure.  To bridge this gap and we use laser scanning confocal microscopy (LSCM) as our routine real space imaging tool. LSCM provides further insight into our network structures as it allows 3D visualization of a sample. Experiments are performed with an in-house Nikon A1R MP laser scanning confocal microscope. The excitation wavelength is 402 nm or 457 nm. As the IPG photoresist shows sufficient autofluorescence at both of these excitation wavelengths, further addition of a dye to the photoresist is not necessary. The sample is imaged in a glass cell filled with toluene, which ensures clear samples by matching the refractive index of the solvent ($n_{toluene}$ = 1.496) and the polymer , ($n_{resist}$ = 1.52). Similar results are obtained using microscope immersion oil as a clearing agent. Typical results are shown in  Figure \ref{Fig7}. The images clearly reveal the  3D network structure and each dielectric rod is visible. Images in Figure \ref{Fig7} a-b show 3D reconstructions of a 35x35x14 $\mu m$$^3$ sub-volume of a sample whereas Figure \ref{Fig7} c shows a cross-section of the structure, parallel to the sample surface at the height of approximately 7 $\mu m$. No signs of overexposure or bleaching of the autofluorescence, neither at the surface nor in the bulk of the sample, are observed. Moreover we can use the 3D reconstruction to numerically calculate the diffraction pattern by 3D Fourier transformation. We note that the result of this procedure are affected by the point spread function of the microscope objective. In Fourier space this translates to a convolution of the scattering function with the point spread function of the objective\cite{deconvolutionreview}. Nevertheless, as long as the structural length scales under study are large compared to the wavelength $a\gg\lambda$, we expect the additional smearing to be relatively small for our high numerical aperture microscope objective (Nikon NA 1.4 100$\times$ oil-immersion objective). Indeed the results obtained by this procedure, Figure \ref{Fig7}  d,  overall compare well to the scattering experiments and to the numerical calculations performed using the original template derived from the seed pattern as reported in ref. \cite{Haberko13}.

\section{Numerical calculations of the scattering function I(q)}
\label{sec:4}
\subsection{Implementation of the numerical calculations}
\label{subsec:4.1}
We model the light scattering properties of our structures in the weak scattering limit using a discrete dipole approximation (DDA) \cite{Draine}. The numerical procedure is implemented as follows: first we prepare a three dimensional binary representation of our structures using a sufficiently densely spaced array. Following that we calculate the 3D Fast Fourier Transform of our dataset and then its modulus squared, which is proportional to the scattered light intensity $I(\vec{q})$ (or the differential scattering cross section) for a scattering vector $\bf{q}=\bf{k}-\bf{k_0}$, where $\bf{k}$ denotes the scattered wave vector and $\bf{k_0}$ the incident wave vector. Since the structures are disordered, the scattering pattern is isotropic : a typical diffraction pattern consists of concentric rings with no distinct Bragg peaks. We thus calculate radial averages of the simulated scattering pattern, as the angular dependence carries no structural information. Note that the (numerical and experimental) diffraction pattern also show the grainy speckle pattern characteristic of scattering from disordered materials \cite{goodman}. The grid density of the DDA calculations has been chosen such that we can accurately reproduce the structure of our metamaterial while keeping the computational load at an acceptable level.  Following these considerations the network structures were reproduced in silico using MatLab \cite{MatLab} with a resolution of one dipole per site every $70$ nm. In order to verify the accuracy of this procedure we also carried out a small number of calculations  doubling the resolution (one dipole site per $35$ nm). We find no significant difference and thus conclude that the accuracy of our procedure is sufficient.

\subsection{Finite size effects}
\label{subsec:4.2}
\begin{figure}\centering
\resizebox{0.5\textwidth}{!}{\includegraphics{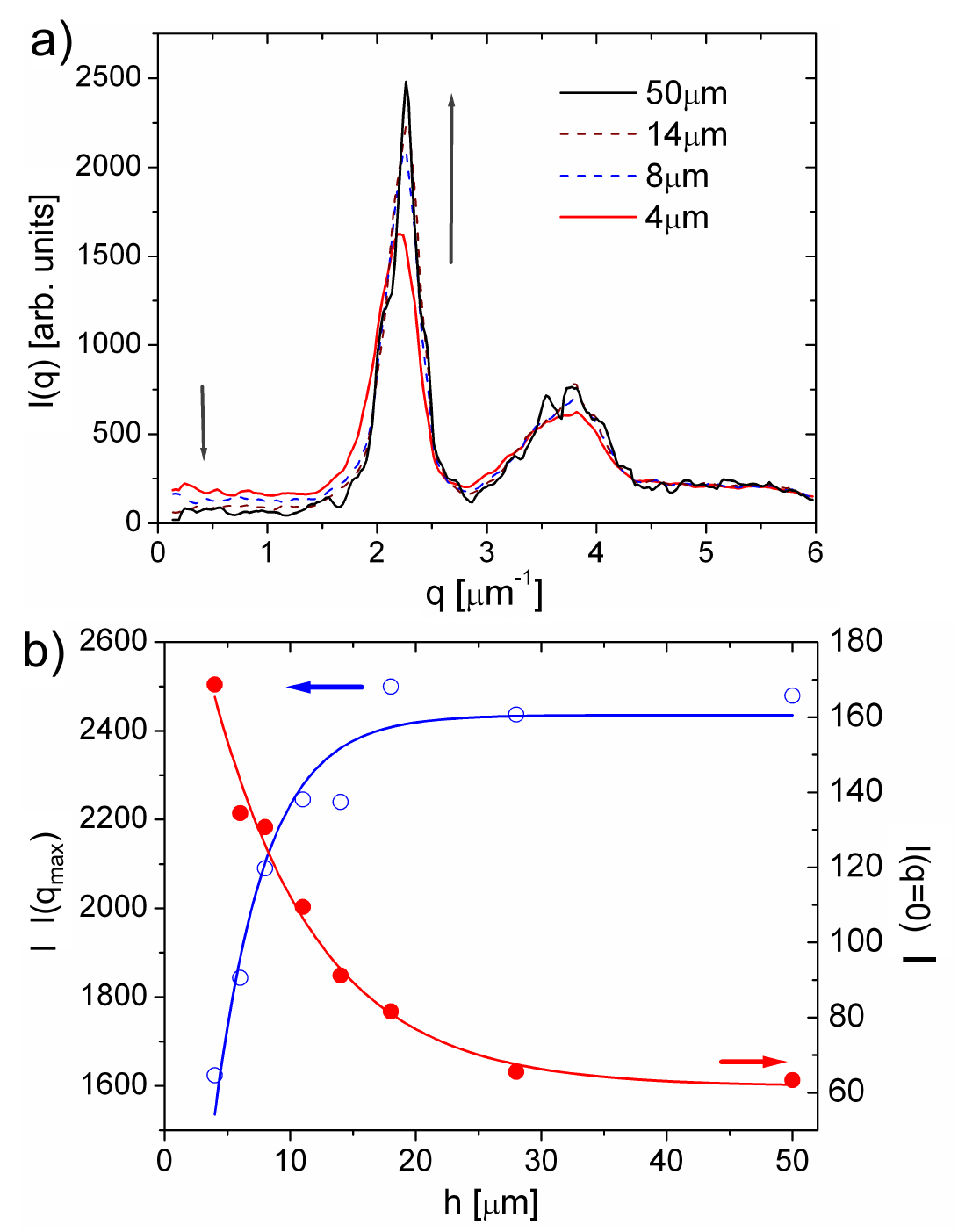}
} 
\caption{(Color online). a) Radially averaged diffraction patterns $I(q)$ derived from computer-generated 3D network structures as a function of height $h$ with $a=3.3 \mu$m (laser pen cross section $0.28 \mu$m$\times 0.84\mu$m) . b) Peak height (left axis) and low $q$ values (values averaged between 0.4 $\mu m^{-1}$ and 1.2 $\mu m^{-1}$, right axis) as a function of height. The footprint of the structures is $50 \times 50 \mu$m.  }
\label{Fig8}      
\end{figure}

In \cite{Haberko13} we demonstrated that the measured and the calculated scattering functions are in good quantitative agreement. We can thus use the numerical model calculations to  investigate systematically how different effects influence the scattering function of the sample. We first address the role of finite size effects. The main limitation in size is the finite height $h \le 12 \mu$m of the structures while the footprint of our structures is comparably large, with a diameter of the order of $\sim 50 \mu$m. Already our first results, ref. \cite{Haberko13}, clearly indicated that the scattering function has not completely evolved for the sample heights considered. On the other hand it is difficult to fabricate very high structures due to other competing effects such as the pyramidal distortions discussed below. Moreover, writing large volume structures is very time consuming. We thus calculate diffraction pattern for different heights $h$, ranging from 4 $\mu m$ to 50 $\mu m$ for structures with a footprint of $50\times50 \mu$m and $a=3.3\mu$m. Figure \ref{Fig8}  a-b shows the radially-averaged $I(q)$. The scattered intensities for each $h$ are normalized by the number of scatterers to facilitate comparison between structures of different height. We observe that the height of the primary peak, located at around $q\simeq$2.2 $\mu m^{-1}$, increases continously with height. It can be noticed, however, that the change in peak height is most dramatic for low structures, whereas for $h\gg a$ the scattering properties asymptotically approach a bulk value. Of particular importance is the behaviour at $q \to 0$  since hyperuniformity should lead to a vanishing scattering function $I(q)$ at small wavenumbers. The latter appears to be a key property of the disordered photonic materials suggested by Florescu and coworkers \cite{Torq09}. In order to assess the low $q$ behavior we analyze the average scattered intensity in the range between $q$ = 0.4 $\mu m^{-1}$ and 1.2 $\mu m^{-1}$ (Figure \ref{Fig8}b). Indeed the low $q$ scattering strongly decreases as the  height of the structures is increased. Asymptotically we do not observe this contribution to tend towards zero. We think the residual scattering even for rather high structures is (at least partially) due to the finite lateral size of the structures. Due to the limited number of points in the seed pattern, extracted from reference \cite{ref15}, we have not yet been able to verify this assumption and we will address the question in a future study using larger systems.  The height dependence of $I(q_{max})$ and $I(q\to0)$ can be fitted with an exponential decay/rise, Figure \ref{Fig8} b. The characteristic lengths obtained by fitting the $h-$dependence are $h_{peak} \sim 1.2 a$ and $h_{zero} \sim 2.5 a$, respectively.

\subsection{Pyramidal distortions}
\begin{figure}\centering
\resizebox{0.4\textwidth}{!}{\includegraphics{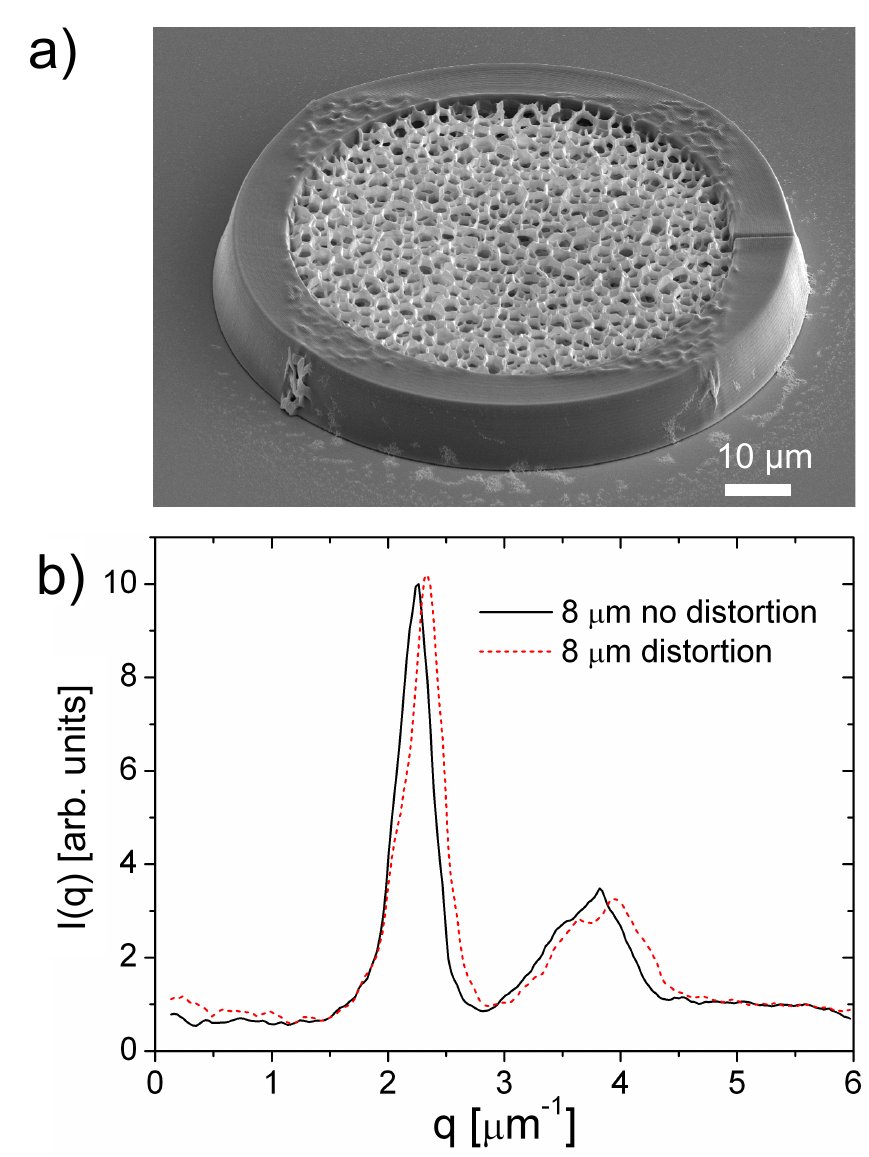}
} 
\caption{(Color online). a.) Scanning electron micrographs (oblique view) of a structure of height $h=12\mu$m. b.) Scattering function $I(q)$ of a computer-generated network structure with and without a pyramidal distortion.}
\label{Fig9}      
\end{figure}

When writing structures higher than two or three basic units ($h \ge 2 a$), we observed a pyramidal distortion, visible in scanning electron micrographs, Figure \ref{Fig9} a). These distortions are mostly likely due to stresses that build up during polymerization and development. At the glass interface the stresses are compensated by interactions with the substrate while at larger height this leads to visible shrinkage. From electron microscopy we estimate a distortion angle of approximately $15^\circ$. In order to assess the influence of such unwanted deformations we calculate the diffraction patterns of both non-distorted samples (which, for the time being, we are not able to fabricate) and the distorted ones. Fortunately we find that the distortion does not significantly affect the scattering function $I(q)$ but only results in a slight shift of $I(q)$ towards larger $q$. This effect is caused by the shrinkage of the upper parts of the structure, slightly decreasing the characteristic distances inside the structure. For a 8 $\mu m$-high sample we estimate, from electron microscopy images, that the linear shrinkage increases from 0 close to the substrate to 7.8$\%$ at the free surface at the top, corresponding to an average shrinkage of 3.9$\%$. From the calculations, Figure \ref{Fig9}, we estimate a shift of $\Delta q \sim 0.08 \mu$m$^{-1}$ corresponding to shrinkage of 3.5$\%$, which is in good agreement with the former value.

\subsection{Influence of the ellipticity of the laser pen}
\begin{figure}\centering
\resizebox{0.43\textwidth}{!}{\includegraphics{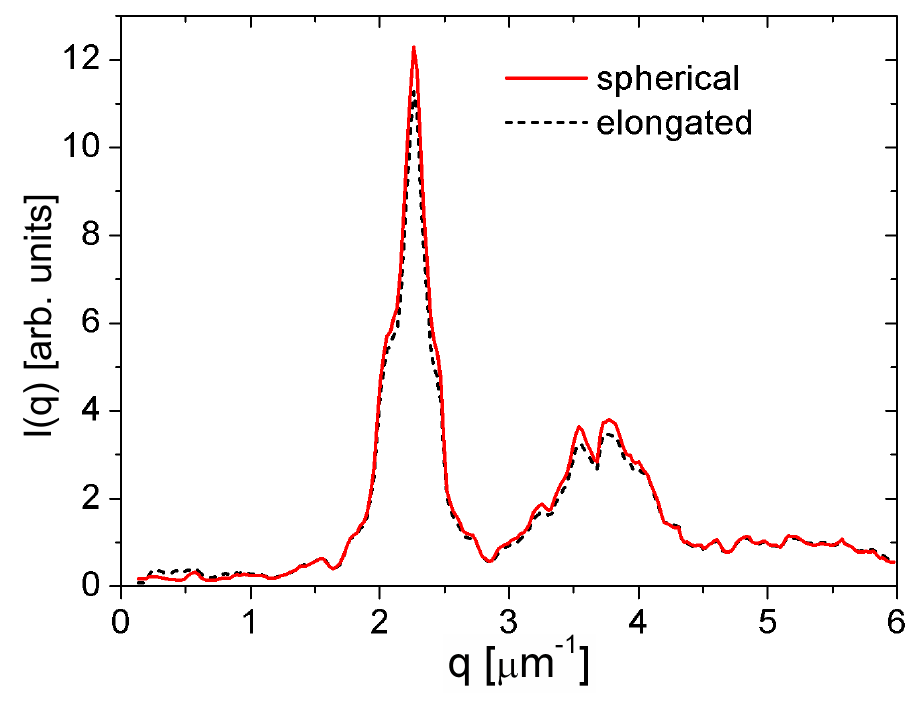}
} 
\caption{(Color online). Scattering function $I(q)$ derived from computer-generated 3D network structures ($a=3.3\mu$m, $h=50\mu$m) using either an ellipsoidal (elongated, $0.28 \mu$m$\times 0.84\mu$m) or spherical (diameter $0.458\mu$m) laser writing pen.}
\label{Fig10}      
\end{figure}

The cross-section and aspect ratio of dielectric rods fabricated by direct laser writing depends on the shape and size of the isointensity surface of the focussed laser spot at the cross-linking threshold of the polymer photoresist \cite{shadedring2}. The isointensity surface is given by the point spread function of the objective for the given illumination conditions. In general it is an ellipsoid, elongated in the direction of propagation of the laser beam. An important consequence of this is that the cross-section of a given rod depends on its orientation in space. All rods perpendicular to the sample surface (or parallel to the beam) have a circular cross-section, whereas those parallel to the surface are elliptic. We have thus examined numerically whether the light scattering properties of a computer-generated structure are affected by this difference in rod shape within the structures.  Starting from the same seed pattern we construct a network either using the elliptical pen shape with an aspect ratio of $2.9$ or a spherical pen. As before we divide the calculated scattering function by the total number of dipoles.  The results in Figure \ref{Fig10}  show that the elliptical pen shape has very little effect on the radially averaged scattering function $I(q)$. The only visible difference is the fact that the peak height is slightly increased for the case of a structure obtained with a spherical pen shape.

\section{Summary and conclusions}
\label{sec:6}
In this article we have presented a detailed study of the different parameters affecting the fabrication of disordered hyperuniform polymeric templates for photonic materials. Expanding on our previous work \cite{Haberko13} we succeeded in reducing all length scales by a factor $5/3$. At the current stage we are able to fabricate routinely polymeric networks starting from a seed pattern with an average point-to-point distance of $a=2 \mu$m and an average rod length of $d\simeq0.8\mu$m in the network after tesselation. Moreover, we have introduced and discussed several characterization tools that allow us to assess the quality of the fabricated polymer structures both in real space and in reciprocal space. A substantial part of this study has been dedicated to the analysis of the various parameters and experimental artefacts that affect the quality of the fabricated structures in practice. The numerical study shows that a minimum height $h\ge 3 a$ of the structure is required for the photonic properties to fully develop. For structures smaller than this finite size effects play an important role. Our numerical studies of the scattering function $I(q)$ show that imperfections due to distortions and the ellipsoidal shape of the writing pen have only a limited impact on the scattering function. Nonetheless further improvements in accuracy might be required when writing larger and higher structures. Also the ellipsoidal shape might adversely affect the spectral properties when a photonic gap opens up at higher refractive index contrast. The pyramidal distortion could be reduced by pre-structuring the substrate with hollow cylindrical holes using classical 2D lithography. Writing the structure inside such voids would allow to connect the outer part of the structure to a solid support that would not deform during development. A similar quality improvement could be obtained by increasing the thickness of the sturdy wall used already now to limit the deformation as shown in Figures \ref{Fig4} and \ref{Fig9} . Both strategies are currently tested in our laboratories. Reducing the aspect ratio of the laser pen on the other side always comes at the expense of reduced writing resolution. Shaping the point spread function of the laser pen with a shaded ring filter is the optimal strategy with minimal loss in resolution \cite{shadedring2,shadedring1}. Design, alignment and characterization are however rather difficult and our current implementation is still experimental (and the commercial version marketed by Nanoscribe  GmbH, Germany, has been withdrawn from their product portfolio in the meantime). Another strategy consists in writing several ellipsoidal voxels side by side thereby reducing the aspect ratio. The latter strategy is reliable and allows reducing the aspect ratio to values close to one. It does however also reduce the actual resolution roughly by a factor equivalent to the aspect ratio of the voxel, so in our case by about a factor 3.

\section{Acknowledgments}
We would like to thank the MPI for polymer research (Mainz) for giving us access to their focused-ion-beam (FIB) instrument and we thank Michael Kappl for help with the experiments. We thank Fr\'ed\'eric Cardinaux for discussions and help preparing the manuscript. JH acknowledges financial support from a Sciex Swiss Research Fellowship No. 10.030. The present project has been financially supported by the National Research fund, Luxembourg (project No. 3093332), the Swiss National Science Foundation (projects 132736 and 128729) and the Adolphe Merkle Foundation.
\newline $^\ast$Corresponding author, email: frank.scheffold@unifr.ch


\begin{thebibliography}{}
\bibitem{dip shaded} The effect of the shaded-ring filter has not yet been evaluated for the  dip-in configuration and we have also not focussed our attention on optimzing its use which may leave some room for further improvements.
\end{thebibliography}
\end{document}